\documentclass[preprint,aps,12pt,showpacs,nofootinbib,tightenlines,amsmath,amssymb]{revtex4}
\usepackage{CJK}
\usepackage{color}
\usepackage{epsfig}

\begin{document}

\title{Two Higgs Bi-doublet Model With Spontaneous P and CP Violation and Decoupling Limit to Two Higgs Doublet Model}
\author{ Jin-Yan Liu}
\author{ Li-Ming Wang}
\author{ Yue-Liang Wu}
\author{Yu-Feng Zhou}

\affiliation{
State Key Laboratory of Theoretical Physics \\
Kavli Institute for Theoretical Physics China \\
Institute of Theoretical Physics,
Chinese Academy of Sciences, Beijing 100190, P.R.China }

\begin{abstract}
  The two Higgs bi-doublet left-right symmetric model (2HBDM) as a simple
  extension of the minimal left-right symmetric model  with a single Higgs bi-doublet is motivated to
  realize  both spontaneous P and CP violation while consistent with the low energy phenomenology without
  significant fine tuning.
  By carefully investigating the Higgs potential of the
  model, we find that sizable CP-violating phases are allowed after the
  spontaneous symmetry breaking.  The mass spectra of the extra scalars in the
  2HBDM are significantly different from the ones in the minimal left-right symmetric model. In
  particular, we demonstrate in  the decoupling limit when the
  right-handed gauge symmetry breaking scale is much higher than the
  electroweak scale,  the 2HBDM decouples into general two
  Higgs doublet model (2HDM) with spontaneous CP violation and has rich induced
  sources of CP violation. We show that in the decoupling limit, it contains extra light Higgs bosons with
  masses around electroweak scale, which can be directly searched at the ongoing
  LHC and future ILC experiments.
\end{abstract}

\pacs{12.60.Fr;12.60.Cn;11.15Ex;11.30Er}

\maketitle
\newpage

\section{Introduction}

The left-right symmetric models\cite{LRM1,LRM2,LRM3} based on the gauge
group $SU(2)_L\times SU(2)_R\times U(1)_{B-L}$ are extensions of the standard
model (SM) motivated by explaining the origin of parity(P) violation and the smallness
of neutrino masses.  In general, it is expected that charge conjugation and parity (CP) violation can also be
realized as a consequence of spontaneous symmetry breaking\cite{TDLee} in this type of
models\cite{LRM4,LRM5,LRM6,Ecker,LRM7}.
One of the extensively studied left-right symmetric models is the minimal left-right symmetric
model which contains two $SU(2)$ triplets and one bi-doublet in the Higgs
sector. Despite its simplicity and success in generating the tiny neutrino
masses, it suffers from a series of constraints in the Higgs and fermion
sector from low energy phenomenology.
It has been shown that in this model the lightest extra Higgs boson has to be
heavier than $\sim 10$ TeV in order to suppress the tree level flavor-changing
neutral current (FCNC) in neutral kaon meson
mixing~\cite{LRM7,FCNC,NSCP1}.
The conditions for minimizing the Higgs potential lead to the observation that
without significant fine-tuning in the potential parameters, the CP phases in
the vacuum expectation values (VEVs) of the Higgs fields are nearly vanishing\cite{VMC1,VMC2,VMC3}.
 In the  minimal left-right symmetric model the Yukawa couplings for both neutral and
charged Higgs bosons are fixed by the quark masses and
Cabbibo-Kobayashi-Maskawa (CKM) matrix, so that all the CP violating phases
are calculable quantities in terms of quark masses and the ratios of the VEVs
of the bi-doublet. It has been shown that in the decoupling limit in which the
vacuum expectation of the right-handed triplet approaches infinity, the model
fails to reproduce the precisely measured weak phase angle $\sin 2\beta$ from
B factories\cite{LRM7}.
%
%
Furthermore, from the VEV see-saw mechanism, the $\beta$ parameters in the
Higgs potential have to be fine-tuned to be 6-7 order of magnitudes smaller
than other model parameters in order to meet the experimental bound on both
light and heavy neutrino masses\cite{VMC3}, if  the right-handed scale
remains in the TeV range which is accessible by the current large hadron
collider (LHC).
Given the above mentioned difficulties in the  minimal left-right symmetric model, one may simply gave up the
spontaneous CP violation in the minimal left-right symmetric model by
considering explicit CP violation in the Higgs potential and/or the Yukawa
sector\cite{Wp2,Wp3,ubiq,LRM10}. However, a detail analysis shown that
little improvement can be achieved in phenomenology.  An alternative treatment
for spontaneous P and CP violation was to introduce mirror particles in a
model based on $[SU(2)\times U(1)]^2$ gauge symmetry\cite{MR-1,MR-2}.

Motivated by the success of generating spontaneous CP violation from the
general two Higgs doublet model (2HDM)~\cite{2HDM1,WW,2HDM2,HW}, an extension
of the minimal left-right symmetric model with two Higgs bi-doublets (2HBDM)
which can break the CP symmetry spontaneously has been
proposed~\cite{WZ1,WZ2}.
In this paper we show how the 2HBDM
can relax the stringent constraints mentioned above for the  minimal left-right symmetric model, and in which
case it can decouple to the 2HDM. It has been shown in \cite{WZ1,WZ2} that such
a simply extended model can be consistent with the low energy phenomenology in
flavor physics. In this work, we shall concentrate on the details of the
generalized Higgs potential and the vacuum minimal conditions, and demonstrate
how such a model can avoid the  fine-tuning problem in generating  sizable CP violating phases.
so that the left-right symmetric
2HBDM with spontaneous P and CP violation could become more realistic at the
TeV-scale. We focus on the mass spectrum of Higgs bosons in the
2HBDM. Different from the minimal model with only one light neutral Higgs
boson similar to the standard model, we shall show that there exist in general
three light neutral Higgs bosons and one pair of light charged Higgs bosons in
the decoupling limit of 2HBDM, which means that the 2HBDM decouples to 2HDM
when $v_R\rightarrow\infty$. Such a feature differs completely from the
 minimal left-right symmetric model.

The paper is organized as follows: In Sec. II, we give an overview of the
problems appearing in the  minimal left-right symmetric model. 
In Sec. III, we present the most general Higgs potential with two
Higgs bi-doublets, and demonstrate in an explicit way why such a
generalization can save the left-right symmetric model from the above mentioned problems arising
in the  minimal left-right symmetric model, and  the possible new physics at TeV scale .  In
Sec. IV, we show that  the 2HBDM can decouple to 2HDM in  the decoupling limit
and then extend the result to  general cases. The conclusions and remarks are
given in the last section.

\section{Overview of  the minimal left-right symmetric model}
The Higgs sector in the minimal model is consisted of one Higgs bi-doublet
and two Higgs triplets:
\begin{eqnarray}
\phi=\left(
       \begin{array}{cc}
         \phi_1^0 & \phi_2^+ \\
         \phi_1^- & \phi_2^0 \\
       \end{array}
     \right)\sim (2,2,0) ,
\end{eqnarray}
\begin{eqnarray}
\Delta_{L}=\left(
           \begin{array}{cc}
             \delta_{L}^+/\sqrt{2} & \delta_{L}^{++}\\
             \delta_{L}^0 & -\delta_{L}^+/\sqrt{2} \\
           \end{array}
         \right)\sim (3,1,2)
\end{eqnarray}
\begin{eqnarray}
\Delta_{R}=\left(
           \begin{array}{cc}
             \delta_{R}^+/\sqrt{2} & \delta_{R}^{++} \\
             \delta_{R}^0 & -\delta_{R}^+/\sqrt{2} \\
           \end{array}
         \right)\sim (1,3,2)
\end{eqnarray}
where the numbers in the brackets denote the quantum number of Higgs
multiplets under the gauge group $SU(2)_L\times SU(2)_R\times
U(1)_{B-L}$. The neutral parts of Higgs fields obtain VEV in such
pattern:
\begin{eqnarray}
\langle\phi\rangle=\left(
                     \begin{array}{cc}
                       k_1/\sqrt{2}  & 0 \\
                       0 & k_2e^{i\theta_2}/\sqrt{2} \\
                     \end{array}
                   \right)\hspace{1cm}
\langle\Delta_{L,R}\rangle=\left( \begin{array}{cc}
                       0  & 0 \\
                       v_{L,R}e^{i\theta_{L,R}}/\sqrt{2} & 0 \\
                     \end{array}
                   \right) . \hspace{1cm}
\end{eqnarray}

And the most general Higgs potential is given by\cite{VMC3}:
\begin{eqnarray}
V_{\phi\Delta}&=&-\mu_1^2Tr(\phi^+\phi)-\mu_2^2[Tr(\tilde\phi\phi^++\tilde\phi^+\phi)]-\mu_3^2\left[Tr(\Delta_L\Delta_L^+)+Tr(\Delta_R\Delta_R^+)\right]\nonumber\\
&+&\lambda_1Tr^2(\phi\phi^+)\!+\!\lambda_2\left[Tr^2(\tilde\phi\phi^+)\!+\!Tr^2(\tilde\phi^+\phi)\right]\! \nonumber \\ & +& \!\lambda_3Tr(\tilde\phi\phi^+)Tr(\tilde\phi^+\phi)\!+\!\lambda_4Tr(\phi\phi^+)Tr(\tilde\phi\phi^+\!+\!\tilde\phi^+\phi)\nonumber\\
&+&\rho_1\left[Tr^2(\Delta_L\Delta_L^+)+Tr^2(\Delta_R\Delta_R^+)\right] \nonumber \\
& + & \rho_2\left[Tr(\Delta_L\Delta_L)Tr(\Delta_L^+\Delta_L^+)+Tr(\Delta_R\Delta_R)Tr(\Delta_R^+\Delta_R^+)\right]\nonumber\\
&+&\rho_3Tr(\Delta_L\Delta_L^+)Tr(\Delta_R\Delta_R^+)+\rho_4\left[Tr(\Delta_L\Delta_L)Tr(\Delta_R^+\Delta_R^+)+Tr(\Delta_L^+\Delta_L^+)Tr(\Delta_R\Delta_R)\right]\nonumber\\
&+&\alpha_1Tr(\phi\phi^+)\left[Tr(\Delta_L\Delta_L^++\Delta_R\Delta_R^+)\right]+\alpha_2Tr(\tilde\phi\phi^++\tilde\phi^+\phi)Tr(\Delta_R\Delta_R^++\Delta_L\Delta_L^+)\nonumber\\
&+&\alpha_3Tr(\phi\phi^+\Delta_L\Delta_L^++\phi^+\phi\Delta_R\Delta_R^+)\nonumber\\
&+&\beta_1Tr(\phi\Delta_R\phi^+\Delta_L^++\phi^+\Delta_L\phi\Delta_R^+)+\beta_2Tr(\tilde\phi\Delta_R\phi^+\Delta_L^++\tilde\phi^+\Delta_L\phi\Delta_R^+)\nonumber\\
&+&\beta_3Tr(\phi\Delta_R\tilde\phi^+\Delta_L^++\phi^+\Delta_L\tilde\phi\Delta_R^+)
\end{eqnarray}
There are three independent vacuum minimal conditions, after
eliminating $\mu_{1,2,3}$ parameters:
\begin{subequations}
\begin{equation}
(2\rho_1-\rho_3)v_Lv_R=\beta_1k_1k_2\cos(\theta_L-\theta_2)
+\beta_2k_1^2\cos\theta_L+\beta_3k_2^2\cos(\theta_L-2\theta_2),
\label{vsw.a}
\end{equation}
\begin{equation}
0=\beta_1k_1k_2\sin(\theta_L-\theta_2)+\beta_2k_1^2\sin\theta_L+\beta_3k_2^2\sin(\theta_L-2\theta_2) , \label{minimal
condition b}
\end{equation}
\begin{equation}
0=k_1k_2[\alpha_3(v_R^2+v_L^2)+(4\lambda_3-8\lambda_2)(k_1^2-k_2^2)] \sin\theta_2 +v_Rv_L\cdot\beta
k^2~\mbox{terms} ,
\label{sh.a}
\end{equation}
\end{subequations}
where $k^2 = k^2_{1}+k^2_{2}$ represents electroweak scale. From Eq.(\ref{vsw.a}) one can
obtain the so-called VEV seesaw relation,
\begin{equation}
\gamma\equiv
\frac{\beta}{\rho}=\frac{v_Lv_R}{k^2},\;\;\;
\label{vsw.b}
\end{equation}
which indicates big gap between $v_L$ and $v_R$ to produce correct small
neutrino mass. If $\rho$ and $\beta$ parameters are within their
normal range, i.e., there is no fine tuning in Higgs sector, $v_R$
has to go up to $10^7$GeV as shown in the literatures\cite{VMC3}. On the
contrary, if $v_R$ is set to several TeV to obtain TeV phenomenology
$\beta$ parameters have to be fine-tuned to $10^{-7}$. The third
equation would lead to severe fine tuning problem and contradict to
phenomenological bounds on neutral Higgs mass. By diagonalizing Higgs
mass matrix, one finds the FCNC violating Higgs mass is
\begin{eqnarray}
M^2_{FCNC}\sim \frac{1}{2}\alpha_3
v_R^2\frac{1}{\sqrt{1-\frac{2k_{1}k_{2}}{k^{2}}}}.
\end{eqnarray}
The lower bound of $M_{FCNC}$ constrained by low energy phenomenology is $10$ TeV.
Thus it is obvious that the third equation is hardly satisfied unless
vacuum phases $\theta_2$ and $\theta_L$ are fine-tuned to very small
values and the model fails to produce right normal-sized vacuum CP
phase. Combining the constraints from the neutral Higgs mass and the FCNC
Higgs mass, one finds immediately that the fine tuning problem is
inevitable in the minimal model: one has to fine-tune $\alpha_3$
when $v_R$ goes up to $10^7$ GeV while keep $M^2_{FCNC}$ is around
$10$ TeV; or else one has to fine-tune $\beta$ when $\alpha_3$
remains normal size.

>From above analysis one sees clearly that in the minimal model there is
severe inconsistence in the Higgs potential for yielding correct
phenomenology. The vacuum minimal condition, neutrino mass and FCNC
bounds contradict with each other, so that one has to make big
concession on the naturalness of the parameters in the Higgs sector,
including the fine-tuned nearly zero vacuum CP phase, losing
elegance and failing in spontaneous CP violation. The fundamental
reason of this self-inconsistence results from the fact that the
fermion-Higgs couplings are too strongly constrained by the
left-right symmetry. This is exactly why we want to add an extra
Higgs bi-doublet to relax the Yukawa sector.

\section{the two Higgs bi-doublet Left-right symmetric model}

We simply add an extra Higgs bi-doublet $\chi$ into the Higgs sector:
\begin{eqnarray}
\chi=\left(
       \begin{array}{cc}
         \chi_1^0 & \chi_2^+ \\
         \chi_1^- & \chi_2^0 \\
       \end{array}
     \right)\sim (2,2,0),
\end{eqnarray}
which has the same
gauge property as $\phi$ in the minimal model. The overall neutral parts of Higgs fields obtain VEV in such
pattern:
\begin{subequations}
\begin{eqnarray}
\langle\phi\rangle=\left(
                     \begin{array}{cc}
                       \kappa_1e^{i\theta_1^p}/\sqrt{2}  & 0 \\
                       0 & \kappa_2e^{i\theta_2^p}/\sqrt{2} \\
                     \end{array}
                   \right),\;\;
\langle\chi\rangle=\left(
                     \begin{array}{cc}
                       \omega_1e^{i\theta_1^c}/\sqrt{2}  & 0 \\
                       0 & \omega_2e^{i\theta_2^c}/\sqrt{2} \\
                     \end{array}
                   \right),
\end{eqnarray}
\begin{eqnarray}
\langle\Delta_{L}\rangle=\left( \begin{array}{cc}
                       0  & 0 \\
                       v_{L}e^{i\theta_L}/\sqrt{2} & 0 \\
                     \end{array}
                   \right),\;\;
\langle\Delta_{R}\rangle=\left( \begin{array}{cc}
                       0  & 0 \\
                       v_Re^{i\theta_R}/\sqrt{2} & 0 \\
                     \end{array}
                   \right).
\end{eqnarray}
\end{subequations}
Note now there are in total six  CP phases in the
vaccum parameters, two of which can be rotated away by
gauge group action. We define here the other four gauge invariant  CP phases,
\begin{equation}
\theta^p=\theta_1^p+\theta_2^p,\;\;
\theta^c=\theta_1^c+\theta_2^c,\;\;
\theta^{pc}=(\theta_1^c-\theta_2^c)-(\theta_1^p-\theta_2^p),\;\;
\theta^{LR}=\theta_L-\theta_R
\label{gphase}
\end{equation}

In our following discussion, we take $\theta_1^p=\theta_R=0$ unless otherwise noted.
Next we will comment shortly on the new features of model structures.

\subsection{New Features of  2HBDM }
With the extra bi-doublet $\chi$, the resulting Lagrangian of 2HBDM has new
features in its structure at tree level, which leads to remarkable difference of
phenomenological descriptions. To facilitate further discussion,
we assume that the vev's satisfy the hierarchy structure  $v_L\ll\kappa_{1,2},\omega_{1,2}\ll v_R$. Also the
Parity P and CP symmetry are assumed.

\paragraph*{{\bf Gauge Sector}}

There is no change on the Fermion-gauge part. The Higgs-gauge
sector is altered with more complicated Higgs-gauge interactions. As a result, the
gauge boson mass after spontaneous symmetry breaking (SSB) is slightly changed. The mass matrices
for charged gauge bosons under basis $\{W_L^+,W_R^+\}$ and for neutral ones under basis
$\{W_L^3,W_R^3,B\}$ are,

\begin{subequations}
\begin{eqnarray}
\tilde{M}^W&=&\frac{g^2}{4}
\left( \begin{array}{cc} v^2+2v_L^2 &
-2 ({ \bf  \kappa_1^{\ast}\kappa_2 + \omega_1^{\ast}\omega_2})   \\
-2({\bf \kappa_1\kappa_2^{\ast} + \omega_1\omega_2^{\ast} }) &
v^2+2v_R^2 \\
\end{array} \right) ,\label{gmass1}
\end{eqnarray}
\begin{eqnarray}
\tilde{M}^0&=&\frac{1}{2}\left( \begin{array}{ccc}
\frac{g^2}{2} (v^2+4v_L^2) &
-\frac{g^2}{2}v^2 & -2gg'v_L^2 \\ -\frac{g^2}{2}v^2 &
\frac{g^2}{2}(v^2+4v_R^2) & -2gg'v_R^2 \\ -2gg'v_L^2 &
-2gg'v_R^2 &
2g'^2(v_L^2+v_R^2)\\ \end{array} \right),\label{gmass2}
\end{eqnarray}
\end{subequations}
with
\begin{subequations}
\begin{eqnarray}
&&g\equiv g_L=g_R \;,
 v^2 \equiv \kappa_1^2
+\kappa_2^2+\omega_1^2+\omega_2^2\; ,
\end{eqnarray}
\begin{eqnarray}
&&{\bf \kappa_1^{\ast}\kappa_2}=\kappa_1\kappa_2e^{i(\theta_2^p-\theta_1^p)}\;,
{\bf \omega_1^{\ast}\omega_2}=\omega_1\omega_2e^{i(\theta_2^c-\theta_1^c)}\;.\label{wm}
\end{eqnarray}
\end{subequations}
where $v\simeq 246$GeV is
the electroweak scale. Following the same procedure in \cite{Duka,RNM},
one can obtain the physical gauge boson mass and the
mixing angles, where for $Z_{1,2}$ gauge bosons nothing changed except for the definition of $v$.
\begin{eqnarray}
\left(\begin{array}{c}
 Z_L \\ Z_R \\ A
\end{array}\right)=
\left(\begin{array}{ccc}
c_W & -s_Wt_W & -t_W\sqrt{c_{2W}} \\
0 & \sqrt{c_{2W}}/c & -t_W \\
s_W & s_W & \sqrt{c_{2W}}
\end{array}\right)
\left(\begin{array}{c}
W_{3L} \\ W_{3R} \\ B
\end{array}\right)\;,
\end{eqnarray}
where $A$ is the photon, and $Z_{L,R}$  are neutral gauge bosons with mixing,
\begin{eqnarray}
\tilde{M}^Z=\frac{g^2}{4}\left(\begin{array}{cc}
 (v^2+4v_L^2)c_W^2 & -\left(v^2(1-t_W^2)-4v_L^2t_W^2\right)/\sqrt{c_{2W}} \\
-\left(v^2(1-t_W^2)-4v_L^2t_W^2\right)/\sqrt{c_{2W}} & 4c_W^2\left(v_R^2+v^2/s_{2W}^2+t_W^4v_L^2\right)/c_{2W}
\end{array}\right)\;.
\end{eqnarray}
The physical $Z_{1,2}$ gauge bosons are defined by,
\begin{eqnarray}
\left(\begin{array}{c} Z_1\\ Z_2 \end{array}\right)=
\left(\begin{array}{cc}\cos\xi  & -\sin\xi   \\ \sin\xi &   \cos\xi  \end{array}\right)
\left(\begin{array}{c}Z_L \\ Z_R\end{array}\right)\;,
\end{eqnarray}
and the physical masses are found to be
\begin{eqnarray}
M_{Z_{1,2}}^2&=&\frac{1}{4}\bigg\{[g^2v^2+2(g^2+g'^2)(v_L^2+v_R^2)]\nonumber\\
&\mp&\sqrt{[g^2v^2+2(g^2+g'^2)(v_L^2+v_R^2)]^2-4g^2(g^2+2g'^2)(v_L^2+v_R^2)v^2}\bigg\}\;,
\end{eqnarray}
with the mixing angle $\zeta$ given by,
\begin{equation}
\sin2\xi=-\frac{g^2v^2\sqrt{\cos2\theta_W}}{2\cos^2\theta_W(M_{Z_2}^2-M_{Z_1}^2)}\;.
\end{equation}
For the physical gauge bosons $W_{1,2}$, they are defined as
\begin{eqnarray}
\left(\begin{array}{c} W_1^+\\ W_2^+ \\ \end{array}\right)=
\left(\begin{array}{cc}\cos\zeta  & -\sin\zeta e^{-i\eta}  \\ \sin\zeta e^{i\eta} &   \cos\zeta  \\\end{array}\right)
\left(\begin{array}{c}W_L^+\\W_R^+\end{array}\right)\;,
\end{eqnarray}
with masses
\begin{equation}
M_{W_{1,2}}^2=\frac{g^2}{4}\left(v^2+v_L^2+v_R^2\mp\frac{v_R^2-v_L^2}{\cos2\zeta}\right)\;.
\end{equation}
It is noted that the difference here is the
mixing angle between $W_1$ and $W_2$, which is replaced as,
\begin{eqnarray}
\zeta\sim\frac{\tan2\zeta}{2}=-\frac{|\kappa_1\kappa_2e^{i(\theta_2^p-\theta_1^p)}
+\omega_1\omega_2e^{i(\theta_2^c-\theta_1^c)}|}{v_R^2-v_L^2}\simeq
-\;r\times\frac{M^2_{W_1}}{M^2_{W_2}},
\end{eqnarray}
where $r=2|\kappa_1\kappa_2
+\omega_1\omega_2e^{i\theta^{pc}}|/v^2$.

One may see from Eqs.(\ref{gmass1},\ref{wm}) that the imaginary part of $W_{1,2}$ mixing is
\begin{equation}
-\mbox{Im}(\tilde{M}^W_{12})=2w_1w_2\cos(\theta_2^p-\theta_1^p)\sin\theta^{pc}
+ 2(\kappa_1\kappa_2+\omega_1\omega_2\cos\theta^{pc}) \sin(\theta_2^p-\theta_1^p)
\label{wmi}
\end{equation}
and the complex phase $\eta$ is
\begin{equation}
\sin\eta= \frac{\mbox{Im}(\tilde{M}_{12}^W)}{\tan2\zeta(v_R^2-v_L^2)} .
\end{equation}
The second term in Eq.(\ref{wmi}) vanishes when $\theta_2^p-\theta_1^p$
is rotated away by gauge symmetry, whereas the first term always  remains  nonzero
since  $\theta^{pc}$ is gauge invariant. This distinguishes 2HBDM
from the minimal left-right symmetric model in which the phase of $W_{1,2}$ mixing can entirely be rotated away.

The physical Higgs-gauge interaction depends on the mixings of
Higgs sector. The SM-like Higgs coupling to gauge bosons in
the minimal left-right symmetric model resembles those in the SM in the limit $\kappa_2\ll
\kappa_1$ and $\kappa_1\ll v_R$, whereas in 2HBDM the
couplings might differ from the SM ones, which is due to its
2HDM nature in the decoupling limit. We will further discuss it in
following sections.

\paragraph*{{\bf Yukawa Sector}}

The general form of quark Yukawa couplings is:
\begin{eqnarray}
  \mathcal{L}_Y & = & - \sum\limits_{i,j}\bar{Q}_{iL}
  \left( (y_q)_{ij}\phi+ (\tilde{y}_q)_{ij}\tilde{\phi}  +
(h_q)_{ij}\chi+ (\tilde{h}_q)_{ij}\tilde{\chi} \right) Q_{jR} ,\end{eqnarray}
which induces the quark mass term after SSB,
\begin{eqnarray}
\label{quarkmass2}
M_u&=&\frac{1}{\sqrt{2}}\left( y_q
\kappa_1+{\tilde{y}}_{q}\kappa_2 e^{i\theta^p_2} +
h_q \omega_1e^{i\theta^c_1}+\tilde h_q\omega_2e^{i\theta^c_2}
\right),\nonumber\\
M_d&=&\frac{1}{\sqrt{2}}\left( y_q
\kappa_2e^{-i\theta^p_2}+{\tilde{y}}_{q}\kappa_1 +
h_q \omega_2e^{-i\theta^c_2}+\tilde h_q\omega_1e^{-i\theta^c_1}
\right). \label{yukawa}
\end{eqnarray}
Parity P symmetry requires
\begin{equation}
y_q=y_q^{\dagger},\;\;\tilde y_q=\tilde y_q^{\dagger},\;\;
h_q=h_q^{\dagger},\;\;\tilde h_q=\tilde h_q^{\dagger},\;\;
\end{equation}
When both P and CP are required to be broken down spontaneously,
all the Yukawa coupling matrices are real symmetric. As there
are in total $6\times 4$
free parameters in $y_q, \tilde y_q$ and $h_q,\tilde h_q$, two
significant
consequences follow:

 (1) The very stringent bound on the minimal
model largely results from the fact that the CKM phases are all
calculable quantities given quark masses, mixing angles and ratio
of vev's, while in 2HBDM, although the relation $V_L^{CKM}=(V_R^{CKM})^{\ast}$
still holds (pseudo-manifest), there are more freedoms in the Yukawa
couplings and exists no direct connection between CKM phases and
other input parameters.

(2) The general form in Eq.(\ref{yukawa}) generates large FCNC
at tree level. The situation gets worse when the mass of FCNC
Higgs is brought down to the EW scale. As shown in our previous
works\cite{WZ1,WZ2}, the tree level FCNC could be suppressed following the
similar treatment in the general 2HDM by considering the
mechanism of approximate global $U(1)$ family symmetry\cite{HW,WW,2HDM2}
\begin{equation}
(u_{i},d_{i})\rightarrow e^{-i\theta_{i}}(u_{i},d_{i})
\end{equation}
which is motivated by the approximate unity of the CKM matrix.
 As a consequence, $y$, $\tilde{y}$, $h$ and $\tilde{h}$ are nearly diagonal matrices.

\paragraph*{{\bf Higgs Sector}}

Based on the general form of the Higgs potential in minimal left-right symmetric model,
we carefully write down the most general form of Higgs potential
for the 2HBDM, which is listed in App.(\ref{potential}). From that
potential, we can obtain the vacuum minimal conditions and find that the tension between
spontaneous CP violation and scale hierarchy is largely relaxed, hence the new model
can generate sizable vacuum phases as the source of CP violation.
Whereas the VEV see-saw problem still leads to the fine tuning on
the $\beta$ parameters. We shall postpone the details to next section.

The extended model contains in total 28 freedoms in the Higgs
sector, including $4+4+2+2$ neutral ones, $2+2+1+1$ pairs of
charged ones, and two pair of doubly charged ones. After spontaneous symmetry breaking, 2
neutral and 2 pairs of charged freedoms would become Goldstone bosons
absorbed into the longitudinal part of gauge vectors, leaving 10
neutral, 4 pairs of charged and 2 pairs of doubly charged
physical Higgs bosons. After carefully studying the Higgs mass
spectrum, we find that in the 2HBDM there exist more than one light Higgs
bosons which are at the electroweak scale.

>From above analysis, it is seen that the 2HBDM improves the
minimal one by introducing more flexible Yukawa couplings, hence
allowing for free CKM phases, as well as by enlarging the Higgs
sector to avoid the inconsistence in the vacuum minimal conditions. As a consequence,
the 2HBDM can be the realistic model with spontaneous CP violation.


\subsection{Generalized vacuum minimal conditions and spontaneous CP violation}

In the 2HBDM, there are ten independent vacuum parameters which correspond to
ten independent vacuum minimal conditions, i.e.,
\begin{equation}
0=\frac{\partial V}{\partial \kappa_1}=\frac{\partial V}{\partial \kappa_2}=\frac{\partial V}{\partial \omega_1}=\frac{\partial V}{\partial \omega_2}=\frac{\partial V}{\partial v_L}
  =\frac{\partial V}{\partial v_R}=\frac{\partial V}{\partial \theta^p_2}=\frac{\partial V}{\partial \theta^c_1}=\frac{\partial V}{\partial \theta^c_2}=\frac{\partial V}{\partial \theta_L}
\end{equation}

Based on the most general form of Higgs potential given in Eq.(\ref{potential}), we can write down the general form
of all the ten vacuum minimal conditions following the same procedure in the minimal model. By eliminating seven mass dimensional parameters $\mu^2$s, we obtain three independent equations,
\begin{subequations}
\begin{eqnarray}
\label{vmca}
(2\rho_1-\rho_3)v_Lv_R&=&\beta_1^pk_1k_2\cos(\theta_L-\theta_2^p)
+\beta_2^pk_1^2\cos\theta_L
+\beta_3^pk_2^2\cos(\theta_L-2\theta_2^p)\nonumber\\
&+&\beta_1^cw_1w_2\cos(\theta_L+\theta_1^c-\theta_2^c)
+\beta_2^cw_1^2\cos(\theta_L+2\theta_1^c)
+\beta_3^cw_2^2\cos(\theta_L-2\theta_2^c)\nonumber\\
&+&\beta_1^{pc}k_2w_1\cos(\theta_L+\theta_1^c-\theta_2^p)
+\beta_2^{pc}k_1w_2\cos(\theta_L-\theta_2^c)
+\beta_3^{pc}k_1w_1\cos(\theta_L+\theta_1^c)\nonumber\\
&+&\beta_4^{pc}k_2w_2\cos(\theta_L-\theta_2^p-\theta_2^c)\;,
\end{eqnarray}
\begin{eqnarray}
\label{vmcb}
0&=&\beta_1^pk_1k_2\sin(\theta_L-\theta_2^p)
+\beta_2^pk_1^2\sin\theta_L
+\beta_3^pk_2^2\sin(\theta_L-2\theta_2^p)\nonumber\\
&+&\beta_1^cw_1w_2\sin(\theta_L+\theta_1^c-\theta_2^c)
+\beta_2^cw_1^2\sin(\theta_L+2\theta_1^c)
+\beta_3^cw_2^2\sin(\theta_L-2\theta_2^c)\nonumber\\
&+&\beta_1^{pc}k_2w_1\sin(\theta_L+\theta_1^c-\theta_2^p)
+\beta_2^{pc}k_1w_2\sin(\theta_L-\theta_2^c)
+\beta_3^{pc}k_1w_1\sin(\theta_L+\theta_1^c)\nonumber\\
&+&\beta_4^{pc}k_2w_2\sin(\theta_L-\theta_2^p-\theta_2^c)\;,
\end{eqnarray}
\begin{eqnarray}
\label{vmcc}
\frac{v_L}{v_R}\;\beta\;O(v^{2})&=&(1+\frac{v_L^2}{v_R^2})[ \kappa_1\kappa_2\sin\theta_2^p\alpha_3^p
+\omega_1\omega_2\sin(\theta_1^c+\theta_2^c)\alpha^c_3\nonumber\\
&+&(\kappa_2\omega_1\sin(\theta_2^p+\theta_1^c)+\kappa_1\omega_2\sin\theta^c_2)\alpha_3^{pc}\nonumber\\
&+&(\kappa_1\omega_1\sin\theta^c_1-\kappa_2\omega_2\sin(\theta^p_2+\theta^c_2))\alpha_4^{pc}]
+\lambda\;O(\frac{(v^{2})^2}{v_R^2})\;.
\end{eqnarray}
\end{subequations}
where $\lambda$ in the third equation stands for a group of
$\lambda$ parameters. From Eqs.(\ref{vmca}) and (\ref{vmcb}), it is noticed that the $\beta$
parameters still need to be fine-tuned to satisfy the neutrino mass bound for fulfilling
a phenomenological model with $v_R$ at TeV scale.  The possible
explanation for the smallness of $\beta$ parameters could be a
softly breaking  $Z_2$ or the approximate $U(1)_{P-Q}$ symmetry imposed on Higgs field, however these arguments inevitably lead to
difficulty in generating correct quark mass hierarchy and quark
 mixings\cite{VMC3} or may violate the $M_{W_2}\!-\!M_N$ relation
  obtained from experimental constraints such as
$0\nu\beta\beta$\cite{WN} or from the K/B meson mixings\cite{LRM7}.
 In this note we ignore the fine-tuned $\beta$'s and  focus on the spontaneous CP violation.
 Eq.(\ref{vmcc}) has the following hierarchy structure,
\begin{equation}
\label{sh.c}
\alpha \gg \lambda\;O(\frac{v^{2}}{v_R^2})\gg \beta \frac{v_L}{v_R}.\;\;\;\;
\mbox{(scale hierarchy)}
\end{equation}
In the minimal model the terms proportional to  $\alpha_{3}^{c}$, $\alpha_{3}^{pc}$ and  $\alpha_{4}^{pc}$ are vanishing.
Only one term $\kappa_{1}\kappa_{2}\sin\theta^{p}_{2}\alpha_{3}^{p}$ exists, which leads to  an extremely small
CP phase angle $\theta^{p}_{2}$ in order to satisfy Eq. (\ref{vmcc}). However, in the 2HBDM, as there are much
more free parameters, the $\alpha$-terms
can cancel among themselves such that the sum of them is of  the order   $\lambda$ terms($O(\frac{v^{2}}{v_R^2})$),
\begin{eqnarray}
\lambda\;O(\frac{(v^{2})^2}{v_R^2})&\simeq& \kappa_1\kappa_2\sin\theta_2^p\alpha_3^p
+\omega_1\omega_2\sin(\theta_1^c+\theta_2^c)\alpha^c_3
+[\kappa_2\omega_1\sin(\theta_2^p+\theta_1^c)\nonumber\\+\kappa_1\omega_2\sin\theta^c_2]\alpha_3^{pc}
&+&[\kappa_1\omega_1\sin\theta^c_1-\kappa_2\omega_2\sin(\theta^p_2+ \theta^c_2)]\alpha_4^{pc}\label{alpha} ,
\end{eqnarray}
and  the sum of all the $\alpha$ terms may cancel with the   $\lambda$-terms and the final results is
of  the order of $\beta$-terms($O(\frac{v_L}{v_R})$). Thus in the 2HBDM, Eqs.(\ref{vmcc}) and (\ref{sh.c})
can be both satisfied with sizable CP violating phases.
The condition can naturally be satisfied provided
$\alpha_3^p \sim \alpha_3^c \sim \alpha_3^{pc} \sim \alpha_4^{pc}$ and
$\kappa_1\sim\kappa_2\sim\omega_1\sim\omega_2$. Hence it is seen that the
2HBDM potential allows sizable vacuum phases
$\theta_2^p,\theta_1^c,\theta_2^{c}$, which generate spontaneous CP violation after
spontaneous symmetry breaking in the gauge, Higgs and Yukawa sectors through gauge boson mixings, Higgs mixings and quark mixings.


\section{Decoupling Limit to 2HDM}

In this section we give the explicit form of Higgs mass matrix and separate
contributions from different symmetry breaking scales, i.e., $v_{L,R}$ and electroweak scale $k$.
We first study the Higgs sector in the so-called special decoupling limit to 2HDM, then extend it to a general decoupling limit to 2HDM.
~\newline

\paragraph*{{\bf The special decoupling limit to 2HDM:}}
let us first consider a special case of Eq.(\ref{alpha}) with the following limit:
\begin{equation}
v_L \;\; \ll  \;\; \kappa_2,\omega_2  \;\; \ll \;\; \kappa_1,\omega_1 \;\; \ll \;\; v_R
\label{2hdma}
\end{equation}
The  reasons to apply Eq.(\ref{2hdma}) include:
(1) The electroweak precision test and neutrino mass require $v_L\ll v$;
(2) $v/ v_R\sim 100$GeV$/1$TeV $\ll 1$ for TeV new physics;
(3) $\kappa_2,\omega_2\ll \kappa_1,\omega_1$ ensures $W_{1,2}$ mixing around
 $10^{-3}$\cite{RNM,WN} or smaller.
Combining  Eq.(\ref{alpha}), an immediate consequence of Eq.(\ref{2hdma}) is
\begin{equation}
|\alpha_4^{pc}|\ll 1
\label{2hdmb}
\end{equation}
Note that in the above limit the gauge invariant phases defined in Eq.(\ref{gphase})
 reduce to (with current choice $\theta_1^p=\theta_R=0$)
\begin{equation}
\theta^{pc}\equiv \theta_1^c,\;\;\;\theta^{LR}\equiv \theta_L
\end{equation}
and the other two phases $\theta^p$ and $\theta^c$ become physically negligible as $\theta_2^p$ and
$\theta_2^c$ compared to $\theta^{pc}$ hardly affect physical processes.

>From the Higgs potential Eq.(\ref{pla}), it is not difficult to check that there is mass
splitting of bi-doublets. The symmetry in the Higgs potential is firstly broken
down to $SU(2)_L$ due to large $v_R$. The Higgs bi-doublets acquire masses
around $v_R$ scale through the $\alpha$-type couplings. While the $\alpha_{1,2}^{p,c,pc}$ terms do not break the global $SU(2)_R$ symmetry
for bi-doublets in the Higgs potential, thus they do not contribute the masses at
$v_R$ scale to bi-doublets. The rest $\alpha_3^{p,c,pc}$ terms
($\alpha_4^{pc}$ is omitted) contribute to bi-doublet mass in the following way:
\begin{equation}
\langle Tr[(X+X^{\dagger})\Delta_R\Delta_R^{\dagger}]\rangle=Tr[(X+X^{\dagger})\cdot \mathcal{P_R}]\;v_R^2
\end{equation}
with
\begin{equation}
\mathcal{P_R}=\frac{1}{2}\left(\begin{array}{cc} 0 & \\ & 1\\\end{array}\right)
\end{equation}
Here the left-right asymmetric operator $\mathcal{P_R}$ brings the residual effect of $SU(2)_R$
symmetry breaking into electroweak sector, resulting in mass splitting among components
of bi-doublets and the scale hierarchy in the vacuum minimal conditions.  Thus it is clearly seen
the inconsistence inside the minimal model as the left-right asymmetric $\alpha_3$ term is simultaneously linked with both spontaneous CP violation and FCNC, i.e., the spontaneous CP violation requires a fine-tuned $\alpha_3$ of order $k^2/v_R^2$ to generate sizable CP asymmetry, while FCNC bound
requires a large mass splitting of order $10$TeV. In the 2HBDM, such a tension is moderated through more flexible vacuum structure and Yukawa couplings. To be more precise, let us define the following structure
\begin{eqnarray}
\phi\equiv (\phi_1,\phi_2),\;\;\; \chi\equiv (\chi_1,\chi_2)\;\;,\label{split}
\end{eqnarray}
with $\phi_1$, $\phi_2$, $\chi_1$ and $\chi_2$ are four doublets. When omitting the mixing between electroweak scale and $\nu_{R}$ scale, we obtain the following Higgs potential after $SU(2)_{R}$ symmetry breaking
\begin{eqnarray}\label{quad}
  V_{\phi,\chi,\langle\Delta_R\rangle}&=&
  -(\mu_1^p)^2Tr[\phi^{\dagger}\phi]-(\mu_1^c)^2Tr[\chi^{\dagger}\chi]
  -(\mu_1^{pc})^2Tr[\phi^{\dagger}\chi+h.c.] \nonumber\\
& + & \frac{\alpha_{1}^{p}v_R^2}{2}Tr[\phi^{\dagger}\phi]
+\frac{\alpha_{1}^{c}v_R^2}{2}Tr[\chi^{\dagger}\chi]
 + \frac{\alpha_{1}^{pc}v_R^2}{2}Tr[\phi^{\dagger}\chi+h.c.]
 \nonumber\\
&& -(\mu_2^p)^2Tr[\tilde{\phi}^{\dagger}\phi]-(\mu_2^c)^2Tr[\tilde{\chi}^{\dagger}\chi]
  -(\mu_2^{pc})^2Tr[\tilde{\phi}^{\dagger}\chi+h.c.] \nonumber\\
&+& \frac{\alpha_{2}^{p}v_R^2}{2}Tr[\tilde{\phi}^{\dagger}\phi]
+\frac{\alpha_{2}^{c}v_R^2}{2}Tr[\tilde{\chi}^{\dagger}\chi]
 + \frac{\alpha_{2}^{pc}v_R^2}{2}Tr[\tilde{\phi}^{\dagger}\chi+h.c.]\nonumber\\
&+&\frac{\alpha_{3}^{p}v_R^2}{2}Tr[\phi^{\dagger}\phi\mathcal{P_R}]+\frac{\alpha_{3}^{c}v_R^2}{2}Tr[\chi^{\dagger}\chi\mathcal{P_R}]
 + \frac{\alpha_{3}^{pc}v_R^2}{2}Tr[(\phi^{\dagger}\chi+h.c.)\mathcal{P_R}]\nonumber\\
&+& (\lambda-terms) + (\beta-terms)
\end{eqnarray}

 In the limit of Eqs.(\ref{2hdma}) and (\ref{2hdmb}),  all the seven $\mu^2$s parameters can be solved from the vacuum minimal conditions in the form
\begin{eqnarray}
\label{mu}
\frac{(\mu_1^p)^{2}}{v_R^2} &\simeq &\frac{\alpha_1^p}{2},\quad
\frac{(\mu_1^c)^{2}}{v_R^2} \simeq \frac{\alpha_1^c}{2},\quad
\frac{(\mu_1^{pc})^{2}}{v_R^2} \simeq \frac{\alpha_1^{pc}}{2},\quad
\frac{(\mu_3)^{2}}{v_R^2}\simeq \frac{\rho_1}{2}, \nonumber\\
\frac{(\mu_2^p)^{2}}{v_R^2} &\simeq & \frac{\alpha_2^p}{2},\quad
\frac{(\mu_2^c)^{2}}{v_R^2} \simeq \frac{\alpha_2^c}{2},\quad
\frac{(\mu_2^{pc})^{2}}{v_R^2}\simeq \frac{\alpha_2^{pc}}{2}.
\end{eqnarray}
where approximation is made by omitting all electroweak scale contributions
from $\lambda$ terms shown in Eq.(\ref{pll}).

With the definition of Eq.(\ref{split}), we arrive at the corresponding quadratic terms for the four doublets $(\phi_1,\chi_1,\phi_2,\chi_2)$,
\begin{eqnarray}\label{quad}
 V^{(2)}_{\phi,\chi,\langle\Delta_R\rangle}&=&
  -(\tilde\mu_1^p)^2\phi_1^{\dagger}\phi_1-(\tilde\mu_1^c)^2\chi_1^{\dagger}\chi_1
-   (\tilde\mu_1^{pc})^2(\phi_1^{\dagger}\chi_1+\chi_1^{\dagger}\phi_1) \nonumber\\
& - & \left[2(\tilde\mu_2^p)^2 \phi_1^T\varepsilon \phi_2
  + 2(\tilde\mu_2^c)^2 \chi_1^T\varepsilon \chi_2
  +(\tilde\mu_2^{pc})^2 \left(\phi_1^T\varepsilon \chi_2 - \phi_2^T\varepsilon \chi_1 \right)\right]+h.c.
 \nonumber\\
&+& \left(\alpha_3^pv_R^2/2-(\tilde\mu_1^p)^2\right)\phi_2^{\dagger}\phi_2
  + \left(\alpha_3^cv_R^2/2-(\tilde\mu_1^c)^2\right)\chi_2^{\dagger}\chi_2 \nonumber\\
&+& \left(\alpha_3^{pc}v_R^2/2-(\tilde\mu_1^{pc})^2\right)(\phi_2^{\dagger}\chi_2+h.c.)
\nonumber\\
&+& (\lambda-terms) + (\beta-terms)
\end{eqnarray}
with
\begin{eqnarray}
\varepsilon=\left(
       \begin{array}{cc}
         0 & 1 \\
         -1 & 0 \\
       \end{array}
     \right)
\end{eqnarray}
where we have redefined the electroweak scale parameters
\begin{equation}
\tilde\mu_i^2=\mu_i^2-\alpha_iv_R^2/2
\end{equation}
which are reasonably small when applying the vacuum minimal conditions Eq.(\ref{mu}) resulted from the limit case Eqs.(\ref{2hdma}) and (\ref{2hdmb}). It is manifest that $\phi_1$ and $\chi_1$ will acquire small masses at the electroweak scale after $SU(2)_L$
symmetry breaking from $\lambda$ terms in Eq.(\ref{pll}), while
$\phi_2$ and $\chi_2$ have masses at the $v_R$ scale from $\alpha_3^{p,c,pc}$ terms.
Note that approximate mass degeneration of  Higgs fields $\phi_2^0,\phi_2^{\pm}$ in doublet
$\phi_2$ and $\chi_2^0,\chi_2^{\pm}$  in doublet $\chi_2$ reveals the fact
that they are not involved in $SU(2)_L$ symmetry breaking.

When omitting the terms concerning the heavy Higgs fields $\phi_{2}$,$\chi_{2}$ and $\Delta_{L,R}$, we yield the Higgs potential for the electroweak symmetry,
\begin{eqnarray}
  V_{\phi_1,\chi_1,\langle\Delta_R\rangle}&=&
  -(\tilde\mu_1^p)^2\phi_1^{\dagger}\phi_1-(\tilde\mu_1^c)^2\chi_1^{\dagger}\chi_1
  - (\tilde\mu_1^{pc})^2(\phi_1^{\dagger}\chi_1+\chi_1^{\dagger}\phi_1) \nonumber\\
& + &\lambda_1^p(\phi_1^{\dagger}\phi_1)^2+\lambda_1^c(\chi_1^{\dagger}\chi_1)^2
  +\lambda_1^{pc}[(\phi_1^{\dagger}\chi_1)^2+(\chi_1^{\dagger}\phi_1)^2]
  +\lambda_2^{pc}(\phi_1^{\dagger}\chi_1)(\chi_1^{\dagger}\phi_1)\nonumber\\
& + & \lambda_3^{pc}(\phi_1^{\dagger}\phi_1)(\phi_1^{\dagger}\chi_1+\chi_1^{\dagger}\phi_1)
 + \lambda_4^{pc}(\chi_1^{\dagger}\chi_1)(\chi_1^{\dagger}\phi_1+\phi_1^{\dagger}\chi_1)
  +\lambda_7^{pc}(\phi_1^{\dagger}\phi_1)(\chi_1^{\dagger}\chi_1)
\end{eqnarray}
which is exactly in the same form (with ten independent terms: three $\mu$ terms and seven $\lambda$ terms) as the potential in the general
2HDM model with spontaneous CP violation\cite{2HDM1,WW}.

It is easy to check that in this limit the electroweak part of the gauge and Yukawa sectors is 2HDM-like.
For the $SU(2)_L$ gauge-Higgs interactions, it reads
\begin{eqnarray}
Tr[(D_{L}\phi)^{\dagger}D_{L}\phi]=(D_{L}\phi_{1})^{\dagger}(D_{L}\phi_{1})+(D_{L}\phi_{2})^{\dagger}(D_{L}\phi_{2}),\nonumber\\
Tr[(D_{L}\chi)^{\dagger}D_{L}\chi]=(D_{L}\chi_{1})^{\dagger}(D_{L}\chi_{1})+(D_{L}\chi_{2})^{\dagger}(D_{L}\chi_{2})
\end{eqnarray}
and for the Yukawa interactions, the quark-Higgs couplings can be written as
\begin{eqnarray}
 \mathcal{L}^{Y}&=&
  \bar{Q}_{L}(y_{q}\phi+\tilde{y}_{q}\tilde{\phi}+h_{q}\chi+\tilde{h}_{q}\tilde{\chi})Q_R \nonumber\\
& =&\bar{Q}_{L}(y_{q}\phi_{1}+h_{q}\chi_{1})Q_{R}^{u}+
\bar{Q}_{L}(\tilde{y}_{q}\tilde{\phi}_{1}+\tilde{h}_{q}\tilde{\chi}_{1})Q_{R}^{d}   \nonumber\\
& + & \bar{Q}_{L}(y_{q}\phi_{2}+h_{q}\chi_{2})Q_{R}^{u}+
\bar{Q}_{L}(\tilde{y}_{q}\tilde{\phi}_{2}+\tilde{h}_{q}\tilde{\chi}_{2})Q_{R}^{d}
\end{eqnarray}
This is why the limit in Eqs.(\ref{2hdma}) and (\ref{2hdmb}) is called the 2HDM
limit. The decoupling rule of 2HBDM to 2HDM is the basic reason why the 2HBDM can be a realistic model with spontaneous P and CP violation.

Let us now check the Higgs mass matrix. In the limit of Eqs.(\ref{2hdma}) and (\ref{2hdmb}), the mixings between Higgs bi-doublets and triplets vanish, and also the mixings between left-hand tiplet $\Delta_L$ and right-hand $\Delta_R$ becomes negligible small. As a consequence, the $12\times 12$ mass matrix of the neutral Higgs bosons splits into $(8\times 8)_{pc} \oplus (2\times 2)_L \oplus (2\times 2)_R $ on the $\{ \phi_{1,2}^0,\chi_{1,2}^0,\delta_L^0,\delta_R^0 \}$ basis, and the $6\times 6$ mass matrix of the charged Higgs bosons splits into $(4\times 4)_{pc} \oplus 1_L \oplus 1_R$  in the $\{ \phi_{1,2}^+,\chi_{1,2}^+,\delta_L^+,\delta_R^+ \}$ basis. Substituting Eq.(\ref{mu})
into the mass matrix, we find that there is a big mass splitting inside the bi-doublets, thus half of the eight freedoms obtain masses at the
$v_R$ scale, while the rest four freedoms remain at the electroweak scale, one of which becomes the Goldstone boson of $SU(2)_L$ symmetry breaking. The same reasoning applies to the charged Higgs sector. The imaginary part of $\delta_R^0$  and $\delta_R^+$ becomes the Goldstone bosons of $SU(2)_R$ symmetry breaking, the real part of $\delta_R^0$ and $\delta_L^0$ become physical Higgs bosons at $v_R$ scale. To conclude, there are three neutral  and one pair of charged Higgs bosons at the electroweak scale in the limit of Eqs.(\ref{2hdma}) and (\ref{2hdmb}). At the $v_R$ scale, there are seven neutral Higgs bosons, and three pairs of charged Higgs bosons among which two come from Higgs bi-doublets and one from $\delta_L^{\pm}$, as well as two pairs of doubly charged Higgs bosons $\delta_{L,R}^{\pm\pm}$. The specific form of Higgs mass  spectrum is listed in App.\ref{spectrum}.

\paragraph*{{\bf General decoupling limit to 2HDM:}}
In the general case, the form of Higgs mass matrix is rather complicated. All the six vev's and four  phases enter the expression.
However, the above analysis on bi-doublet mass splitting still holds, which means that the heavy freedoms at the $v_R$ scale in two Higgs bi-doublets
are all dominated by the explicit $SU(2)_R$ symmetry breaking terms $\alpha_3^{p,c,pc}$.

It is rather tedious to write down the general form for the mass matrix, but we have carefully checked and confirmed that in the general case without imposing the special limit given in Eqs.(\ref{2hdma}) and (\ref{2hdmb}), there are still three neutral and one pair of charged Higgs bosons at the electroweak scale as long as the $SU(2)_R$ symmetry breaking scale $v_R$ is taken to be much higher than the electroweak scale, i.e.,
\begin{equation}
v_R \gg \kappa_1,\, \kappa_2,\, w_1,\, w_2,\, v_L
\label{GDL}
\end{equation}
which may be regarded as the general decoupling limit for 2HBDM approaching to 2HDM-like. It is also found that the mixings of Higgs sector have the same pattern as that described in App.{\ref{spectrum}}.

From the above analysis, we arrive at the conclusion that the 2HBDM will degenerate to the 2HDM in a general decoupling limit Eq.(\ref{GDL}). The main difference is that in the general decoupling limit the electroweak sector is separated from the right-hand sector associated with $v_R$ scale in a much more complicated way.

The explicit structures of the mass matrices for physical Higgs bosons are given as follows with different scales
\begin{eqnarray}
M^0=\left(
\begin{array}{cccc}
M_h^0 & v^2 & vv_R & vv_L \\
v^2 & M^0_H & vv_R  & vv_L\\
v_Rv & v_Rv & M^0_R  & v_Rv_L\\
v_Lv  & v_Lv & v_Lv_R & M^0_L \\
\end{array}
\right),\;\;
M^{\pm}=\left(
\begin{array}{ccc}
M_h^{\pm} & v^2 & vv_L\\
v^2 & M^{\pm}_H & vv_L\\
v_Lv & v_Lv & M^{\pm}_L \\
\end{array}
\right).
\label{ncm}
\end{eqnarray}
with $M_h^0$ a $3 \times 3$ mass matrix, $M_H^0$  a $4 \times 4$ mass matrix and $M_H^{\pm}$ a $2 \times 2$ mass matrix, and
\begin{eqnarray}
M_H^0&=&
\left(
\begin{array}{cc}
M_H^{0R} & v^2 \\
v^2 & M_H^{0I} \\
\end{array}
\right),\;\;\;
M_L^0=\frac{(\rho_3-2\rho_1)v_R^2}{2}
\left(\begin{array}{cc}
1 \\
 & 1\\
\end{array}\right),\nonumber\\
M_R^0&=&2\rho_1v_R^2\;,\hspace{1.9cm}
M_L^{\pm}=\frac{(\rho_3-2\rho_1)v_R^2}{2} + \alpha v^2\;.
\end{eqnarray}
Thus the neutral Higgs mass matrix $M^0$ in Eq.(\ref{ncm}) is a $10 \times 10$ matrix, and $h^0$($h^{\pm}$)
and $H^0$($H^{\pm}$) stand for the (nearly physical) Higgs bosons from the combination of bi-doublets with
 mass scales $k$ and $v_R$, respectively.
$M_h^0$ is $3\times3$ matrix with elements of order  $k^2$, and $M_h^{\pm}$ is also of order $k^2$, while
$M_H^{0R}$(real part), $M_H^{0I}$(imaginary part) and $M_H^{\pm}$ matrix elements are all of order $v_R^2$.

The Goldstone bosons are defined as
\begin{eqnarray}
\tilde{G}_L^0&=& \mbox{Im}\left({\bf \kappa_1}\phi_1^{0*}+{\bf \kappa_2^*}\phi_2^{0}+{\bf \omega_1}\chi_1^{0*}+{\bf \omega_2^*}\chi_2^{0}+2{\bf v_L^*}\delta_L^0\right)/\sqrt{v^2+4v_L^2}\;,\nonumber\\
\tilde{G}_R^0&=& \mbox{Im} \left({\bf \kappa_1^*}\phi_1^{0}+{\bf \kappa_2}\phi_2^{0*}+{\bf \omega_1^*}\chi_1^{0}+{\bf \omega_2}\chi_2^{0*}+2{\bf v_R^*}\delta_R^0\right)/\sqrt{v^2+4v_R^2}\;,\nonumber\\
\tilde{G}_L^+&=&\left({-\bf \kappa_1}\phi_1^++{\bf \kappa_2^*}\phi_2^+-{\bf \omega_1}\chi_1^++{\bf \omega_2^*}\chi_2^++\sqrt{2}{\bf v_L^*}\delta_L^+\right)/\sqrt{v^2+2v_L^2}\;,\nonumber\\
\tilde{G}_R^+&=&\left(-{\bf \kappa_1^*}\phi_2^++{\bf \kappa_2}\phi_1^+-{\bf \omega_1^*}\chi_2^++{\bf \omega_2}\chi_1^++\sqrt{2}{\bf v_R^*}\delta_R^+\right)/\sqrt{v^2+2v_R^2}\;.
\end{eqnarray}
which have been extracted from the mass matrix.  Note that the real directions of the neutral and charged Goldstone bosons in the general case correspond to the combination of $\tilde{G}_{L,R}^0$ and $\tilde{G}_{L,R}^{\pm}$, respectively.

\paragraph*{{\bf Low energy phenomenology:}}
In our previous works\cite{WZ1,WZ2}, we have shown the low energy phenomenological constraints
and demonstrated that the mentioned stringent phenomenological constraints on the minimal model from neutral
meson mixings can be significantly relaxed. In particular, it has been shown that the right-handed gauge boson
mass can be as low as 600 GeV with the charged Higgs mass around 200 GeV.
The FCNC will not impose severe constraints on the neural Higgs mass, provided small off-diagonal
Yukawa couplings via the mechanism of approximate global $U(1)$ family symmetry\cite{HW,2HDM1,WW}.
We have also analyzed the mass difference $\Delta m_K$ and
indirect CP violation $\epsilon_K$ in the neutral $K$ system and observed that the right-handed
gauge boson contributions to the mass difference $\Delta m_K$ can be
opposite to that from the charged Higgs boson
and a cancelation between the two contributions is possible in a
large parameter space. The suppression of right-handed gauge boson contributions to the
indirect CP violation $\epsilon_K$ has been found to occur naturally.  As
a consequence, a light right-handed gauge boson around the current
experimental low bound is allowed. For the neutral B meson system, the mass difference $\Delta m_B$
and the time dependent CP asymmetry in $B^0\to J/\Psi K_S$ decay have been
found to be consistently characterized in the 2HBDM with spontaneous P and CP violation, which is unlike the minimal model with only one
Higgs bi-doublet.

\section{Conclusion}

We have discussed the 2HBDM with spontaneous P and CP violation as a simple extension of the minimal left-right symmetric model by adding an extra bi-doublet. It has been shown that such an extended 2HBDM can solve the inconsistency between the vacuum minimal
conditions on spontaneous CP violation and the low energy phenomenological bounds on the FCNC Higgs
mass. It has been found that the 2HBDM can relax the quark Yukawa sector which is
strictly constrained by left-right symmetry in the minimal one.

In particular, we have demonstrated the existence of a general decoupling limit in the 2HBDM, which states that as long as the $SU(2)_R$ symmetry breaking scale caused by the $SU(2)_R$ triplet Higgs is much higher than the electroweak symmetry breaking scale, the 2HBDM will degenerate to the 2HDM-like, which apparently differs from the minimal model with or without spontaneous CP violation. As a consequence, the Higgs mass spectra in the 2HBDM have been obtained with reasonable approximation, where the three neutral and one pair of charged Higgs become naturally light Higgs bosons with masses at the electroweak scale and may be explored at LHC and ILC colliders\cite{Bao:2010sz,Bao:2009sa}, and the sources of CP violation in the 2HBDM also get much richer and may show up in low energy processes such as rare $B$ decays\cite{Bao:2008hd,Wu:2007eb,Wu:2006ur,Su:2006uy}.

As the 2HBDM decouples to the 2HDM in the decoupling limit, it can evade the stringent constraints from the neutral meson mixing and make the allowed mass of right-handed gauge boson to be closing to the current direct experimental search bound.  It is expected that the
new physics particles in the 2HBDM can directly be searched in upcoming LHC and future ILC experiments.

\vspace*{0.5cm}

\vspace{1 cm}

\centerline{{\bf Acknowledgement}}

\vspace{20 pt}

This work is supported in part by the National Basic Research
Program of China (973 Program) under Grants No. 2010CB833000; the National
Nature Science Foundation of China (NSFC) under Grants No. 10975170,
No. 10821504 and No. 10905084; and the Project of Knowledge Innovation Program
(PKIP) of the Chinese Academy of Science.  JYL is grateful to  Yun-Jie Huo for helpful discussions.

\newpage

\appendix

\section{The most general Higgs potential in 2HBDM}

With simply adding a Higgs bi-doublet $\chi$, the Higgs potential becomes much
more complicated than the minimal model with a single Higgs bi-doublet. Except including an identical copy
of $\chi$ coupling to triplets $\Delta_{L,R}$, there are also the
mixing terms between two Higgs bi-doublets $\phi$ and $\chi$. The general form of Higgs potential may be written as follows
\begin{subequations}
\label{potential}
\begin{equation}
V_{\phi,\chi,\Delta_L,\Delta_R}=V_{\mu}+V_{\alpha}+V_{\rho}+V_{\lambda}+V_{\beta},
\end{equation}
with
\begin{eqnarray}
  V_{\mu}&=&-(\mu_1^p)^2Tr[\phi^{\dagger}\phi]
-(\mu_1^c)^2Tr[\chi^{\dagger}\chi]\nonumber\\
&-&(\mu^{pc}_1)^2Tr[\phi^+\chi+\chi^+\phi]\nonumber\\
&-&(\mu_2^p)^2Tr[\tilde{\phi}\phi^{\dagger}+\tilde{\phi}^{\dagger}
\phi]
-(\mu_2^c)^2Tr[\tilde{\chi}\chi^{\dagger}+\tilde{\chi}^{\dagger}
\chi]\nonumber\\
&-&(\mu^{pc}_2)^2Tr[\tilde\phi\chi^++\tilde\phi^+\chi]\nonumber\\
&-&\mu_3^2 Tr [\Delta_L \Delta_L^{\dagger}  +
\Delta_R \Delta_R^{\dagger}], \label{plm}
\end{eqnarray}
\begin{eqnarray}
  V_{\alpha}&=&
\alpha_1^p Tr \left[ \phi \phi^{\dagger} \right] Tr \left[
\Delta_L \Delta_L^{\dagger} + \Delta_R \Delta_R^{\dagger}
\right]
+ \alpha_1^c Tr \left[ \chi \chi^{\dagger} \right] Tr \left[
\Delta_L \Delta_L^{\dagger} + \Delta_R \Delta_R^{\dagger}
\right]\nonumber\\
&+&\alpha^{pc}_1Tr\left[\phi^{\dagger}\chi+\chi^{\dagger}\phi\right]Tr\left[\Delta_L\Delta_L^{\dagger}+\Delta_R\Delta_R^{\dagger}\right]\nonumber\\
&+&\alpha_2^p Tr \left[ \phi^{\dagger} \tilde{\phi} + \phi
\tilde{\phi}^{\dagger}\right]
Tr \left[ \Delta_L \Delta_L^{\dagger} + \Delta_R
\Delta_R^{\dagger} \right]
+\alpha_2^c Tr \left[ \chi^{\dagger} \tilde{\chi} + \chi
\tilde{\chi}^{\dagger}\right]
Tr \left[ \Delta_L \Delta_L^{\dagger} + \Delta_R
\Delta_R^{\dagger} \right]\nonumber\\
&+&\alpha^{pc}_2Tr(\tilde\phi\chi^{\dagger}+\tilde\phi^{\dagger}\chi)Tr(\Delta_L\Delta_L^{\dagger}+\Delta_R\Delta_R^{\dagger})\nonumber\\
&+& \alpha_3^p Tr \left[ \phi \phi^{\dagger}(\Delta_L
\Delta_L^{\dagger} + \Delta_R
\Delta_R^{\dagger} ) \right]
+ \alpha_3^c Tr \left[ \chi \chi^{\dagger} (\Delta_L
\Delta_L^{\dagger} + \Delta_R
\Delta_R^{\dagger} ) \right] \nonumber \\
&+&\alpha^{pc}_3Tr\left[\left(\phi\chi^{\dagger}+\chi\phi^{\dagger}\right)\Delta_L\Delta_L^{\dagger}
+\left(\phi^{\dagger}\chi+\chi^{\dagger}\phi\right)\Delta_R\Delta_R^{\dagger}\right]\nonumber\\
&+&\alpha^{pc}_4Tr\left[\left(\tilde\phi\chi^{\dagger}+\chi\tilde\phi^{\dagger}\right)\Delta_L\Delta_L^{\dagger}
+\left(\chi^{\dagger}\tilde\phi+\tilde\phi^{\dagger}\chi\right)\Delta_R\Delta_R^{\dagger}\right], \label{pla}
\end{eqnarray}
\begin{eqnarray}
V_{\lambda}&=&\lambda_1^p Tr ^2 \left[ \phi \phi^{\dagger}
\right]
+ \lambda_2^p \left( Tr ^2\left[ \tilde{\phi} \phi^{\dagger}
\right] + Tr ^2\left[ \tilde{\phi}^{\dagger} \phi \right]
\right)
+ \lambda_3^p Tr \left[ \tilde{\phi} \phi^{\dagger} \right] Tr
\left[
\tilde{\phi}^{\dagger} \phi \right]  \nonumber \\
&+& \lambda_4^p Tr \left[ \phi \phi^{\dagger} \right] \left( Tr
\left[
\tilde{\phi} \phi^{\dagger} \right] + Tr \left[
\tilde{\phi}^{\dagger} \phi
\right] \right) \nonumber\\
&+&\lambda_1^c  Tr ^2 \left[ \chi \chi^{\dagger} \right]
+ \lambda_2^c \left( Tr ^2\left[ \tilde{\chi} \chi^{\dagger}
\right] + Tr ^2\left[ \tilde{\chi}^{\dagger} \chi \right]
\right)
+ \lambda_3^c Tr \left[ \tilde{\chi} \chi^{\dagger} \right] Tr
\left[
\tilde{\chi}^{\dagger} \chi \right]  \nonumber \\
&+& \lambda_4^c Tr \left[ \chi \chi^{\dagger} \right] \left( Tr
\left[
\tilde{\chi} \chi^{\dagger} \right] + Tr \left[
\tilde{\chi}^{\dagger} \chi
\right] \right)\nonumber\\
&+&\lambda^{pc}_1\left(Tr^2\left[\phi^{\dagger}\chi\right]+Tr^2\left[\chi^{\dagger}\phi\right]\right)
+\lambda^{pc}_2Tr\left[\phi^{\dagger}\chi\right]Tr\left[\chi^{\dagger}\phi\right]\nonumber\\
&+&\lambda^{pc}_{3}Tr\left[\phi\phi^{\dagger}\right]Tr\left[\phi^{\dagger}\chi+\chi^{\dagger}\phi\right]
+\lambda^{pc}_{4}Tr\left[\chi\chi^{\dagger}\right]Tr\left[\phi^{\dagger}\chi+\chi^{\dagger}\phi\right]\nonumber\\
&+&\lambda^{pc}_{5}\left(Tr^2\left[\tilde\phi\chi^{\dagger}\right]+Tr^2\left[\tilde\phi^{\dagger}\chi\right]\right)
+\lambda^{pc}_{6}Tr\left[\tilde\phi\chi^{\dagger}\right]Tr\left[\tilde\phi^{\dagger}\chi\right]\nonumber\\
&+&\lambda^{pc}_{7}Tr\left[\phi\phi^{\dagger}\right]Tr\left[\chi\chi^{\dagger}\right]\nonumber\\
&+&\lambda^{pc}_{8}Tr\left[\phi\phi^{\dagger}\right]Tr\left[\tilde\chi\chi^{\dagger}+\tilde\chi^{\dagger}\chi\right]
+\lambda^{pc}_{9}Tr\left[\phi\phi^{\dagger}\right]Tr\left[\tilde\phi\chi^{\dagger}+\tilde\phi^{\dagger}\chi\right]\nonumber\\
&+&\lambda^{pc}_{10}Tr\left[\chi\chi^{\dagger}\right]Tr\left[\tilde\phi\phi^{\dagger}+\tilde\phi^{\dagger}\phi\right]
+\lambda^{pc}_{11}Tr\left[\chi\chi^{\dagger}\right]Tr\left[\tilde\chi\phi^{\dagger}+\tilde\chi^{\dagger}\phi\right]\nonumber\\
&+&\lambda^{pc}_{12,13}Tr\left[\tilde\phi\phi^{\dagger}\pm\tilde\phi^{\dagger}\phi\right]Tr\left[\tilde\phi^{\dagger}\chi\pm\tilde\phi\chi^{\dagger}\right]
+\lambda^{pc}_{14,15}Tr\left[\tilde\chi\chi^{\dagger}\pm\tilde\chi^{\dagger}\chi\right]Tr\left[\tilde\phi^{\dagger}\chi\pm\tilde\phi\chi^{\dagger}\right]\nonumber\\
&+&\lambda^{pc}_{16,17}Tr\left[\phi\chi^{\dagger}\pm\chi\phi^{\dagger}\right]Tr\left[\tilde\phi^{\dagger}\phi\pm\tilde\phi\phi^{\dagger}\right]
+\lambda^{pc}_{18,19}Tr\left[\phi\chi^{\dagger}\pm\chi\phi^{\dagger}\right]Tr\left[\tilde\chi^{\dagger}\chi\pm\tilde\chi\chi^{\dagger}\right]\nonumber\\
&+&\lambda^{pc}_{20,21}Tr\left[\phi\chi^{\dagger}\pm\chi\phi^{\dagger}\right]Tr\left[\tilde\phi^{\dagger}\chi\pm\tilde\phi\chi^{\dagger}\right]\nonumber\\
&+&\lambda^{pc}_{22,23}Tr\left[\tilde\phi^{\dagger}\phi\pm\tilde\phi\phi^{\dagger}\right]Tr\left[\tilde\chi\chi^{\dagger}\pm\tilde\chi^{\dagger}\chi\right], \label{pll}
\end{eqnarray}
\begin{eqnarray}
V_{\beta}&=&
\beta_1^p Tr \left[ \phi \Delta_R \phi^{\dagger}
\Delta_L^{\dagger}
+\phi^{\dagger} \Delta_L \phi \Delta_R^{\dagger}\right]
+\beta_1^c Tr \left[ \chi \Delta_R \chi^{\dagger}
\Delta_L^{\dagger}
+\chi^{\dagger} \Delta_L \chi \Delta_R^{\dagger}
\right]\nonumber\\
&+&\beta^{pc}_1Tr\left[\phi\Delta_R\chi^{\dagger}\Delta_L^{\dagger}+\phi^+\Delta_L\chi\Delta_R^{\dagger}\right]\nonumber\\
&+& \beta_2^p Tr \left[ \tilde{\phi} \Delta_R \phi^{\dagger}
\Delta_L^{\dagger}+  {\tilde{\phi}}^{\dagger} \Delta_L \phi
\Delta_R^{\dagger} \right]
+\beta_2^c  Tr \left[ \tilde{\chi} \Delta_R \chi^{\dagger}
\Delta_L^{\dagger} +{\tilde{\chi}}^{\dagger} \Delta_L \chi
\Delta_R^{\dagger} \right]  \nonumber \\
&+&\beta^{pc}_2Tr\left[\chi\Delta_R\phi^{\dagger}\Delta_L^{\dagger}+\chi^{\dagger}\Delta_L\phi\Delta_R^{\dagger}\right]\nonumber\\
&+& \beta_3^p Tr \left[ \phi \Delta_R {\tilde{\phi}}^{\dagger}
\Delta_L^{\dagger} + \phi^{\dagger} \Delta_L \tilde{\phi}
\Delta_R^{\dagger} \right]
+ \beta_3^c Tr \left[ \chi \Delta_R {\tilde{\chi}}^{\dagger}
\Delta_L^{\dagger} + \chi^{\dagger} \Delta_L \tilde{\chi}
\Delta_R^{\dagger} \right]\nonumber\\
&+&\beta^{pc}_3Tr\left[\tilde\phi\Delta_R\chi^{\dagger}\Delta_L^{\dagger}+\tilde\phi^{\dagger}\Delta_L\chi\Delta_R^{\dagger}\right]
+\beta^{pc}_{4}Tr\left[\phi\Delta_R\tilde\chi^{\dagger}\Delta_L^{\dagger}+\phi^{\dagger}\Delta_L\tilde\chi\Delta_R^{\dagger}\right], \label{plb}
\end{eqnarray}
\begin{eqnarray}
  V_{\rho}&=&
\rho_1 \left(  Tr ^2 \left[ \Delta_L \Delta_L^{\dagger} \right]
+Tr ^2\left[ \Delta_R \Delta_R^{\dagger} \right]
\right)\nonumber\\
&+&\rho_2 \left( Tr \left[ \Delta_L \Delta_L \right] Tr \left[
\Delta_L^{\dagger} \Delta_L^{\dagger} \right] + Tr \left[
\Delta_R \Delta_R
\right] Tr \left[ \Delta_R^{\dagger} \Delta_R^{\dagger}
\right]\right )\nonumber\\
&+& \rho_3 Tr \left[ \Delta_L \Delta_L^{\dagger} \right] Tr
\left[
\Delta_R \Delta_R^{\dagger} \right] \nonumber\\
&+&\rho_4 \left( Tr \left[ \Delta_L \Delta_L \right] Tr \left[
\Delta_R^{\dagger} \Delta_R^{\dagger} \right] + Tr \left[
\Delta_L^{\dagger}
\Delta_L^{\dagger} \right] Tr \left[ \Delta_R \Delta_R \right]
\right). \label{plr}
\end{eqnarray}
\end{subequations}
where upper indices $p's$ denote the terms relevant to $\phi$, $c's$
denote those relevant to $\chi$ and $pc's$ denote those relevant to
both $\phi$ and $\chi$.

\section{Higgs mass spectra in the decoupling limit}\label{spectrum}

Here we explicitly list the Higgs mass spectra in the 2HBDM with approximation by omitting the terms $O(\frac{v_L}{k})$,
$O(\frac{k}{v_R})$ and $O(\frac{\kappa_2,\omega_2}{\kappa_1,\omega_1})$ and their higher orders.
\begin{itemize}
\item Light neutral Higgs ($h^0_1,h^0_2,h^0_3$) at the electroweak scale:

\begin{eqnarray}
(M_{h^0}^{11})^{2}&\simeq&2\lambda_1^p\kappa^2+2\lambda_1^{pc}\omega^2\cos^2\theta_1^c+2\lambda_3^{pc}\kappa\omega\cos\theta,\nonumber\\
(M_{h^0}^{12})^{2}&\simeq&-2\lambda_1^{pc}\kappa\omega\sin^2\theta_1^c+\lambda_2^{pc}\kappa\omega+\lambda_3^{pc}\kappa\omega\cos
+\lambda_4^{pc}\omega^2\cos\theta+\lambda_7^{pc}\kappa\omega,\nonumber\\
(M_{h^0}^{22})^{2}&\simeq&2\lambda_1^c\omega^2+2\lambda_1^{pc}\kappa^2\cos^2\theta_1^c+2\lambda_4^{pc}\kappa\omega\cos\theta,\nonumber\\
(M_{h^0}^{13})^{2}&\simeq&-v(2\lambda_1^{pc}\omega\cos\theta+\lambda_3^{pc}\kappa),\nonumber\\
(M_{h^0}^{23})^{2}&\simeq&-v(2\lambda_1^{pc}\kappa\cos\theta+\lambda_4^{pc}\omega),\nonumber\\
(M_{h^0}^{33})^{2}&\simeq&2\lambda_1^{pc}v^2\sin^2\theta_1^c.
\end{eqnarray}
with
\begin{eqnarray}
\phi_1^0&=&h_1^0-i \sin\beta\;h_2^0+i\cos\beta\; \tilde G_1^0,\nonumber\\
\phi_2^0&=&e^{i\theta}h_3^0-e^{-i\theta}\sin\beta\;h_2^0-e^{-i\theta}\cos\beta\;\tilde G_1^0.
\end{eqnarray}
where  $\theta\equiv \theta_1^c$, $\kappa\equiv \kappa_1$, $\omega\equiv \omega_1$, $v^2=\kappa^2+\omega^2$ and $\tan\beta=\omega/\kappa$. $\tilde G_1^0$ absorbed by $Z_1$ is the neutral Goldstone of $SU(2)_L$ symmetry breaking.

\item Light charged Higgs ($h^{\pm}$) at the electroweak scale:

\begin{eqnarray}
(M_{h^{\pm}})^{2}=\frac{1}{2}(2\lambda_1^{pc}-\lambda_2^{pc})v^2\;.
\end{eqnarray}
with
\begin{eqnarray}
\phi_1^+&=&-\sin\beta e^{-i\theta}\;h^++\cos\beta\;G_L^+,\nonumber\\
\chi_1^+&=&\cos\beta \;h^++\sin\beta e^{i\theta}\;G_L^+.
\end{eqnarray}
where $G_L^+$ absorbed by $W_L$ is the charged Goldstone of $SU(2)_L$ symmetry breaking.

\item Heavy doublets ($H_1,H_2$) on the basis of $\{\phi_2,\chi_2\}$ at the $v_R$ scale:

\begin{eqnarray}
 (M^H_{1,2})^{2}=\frac{1}{2} \left(\alpha_3^p+\alpha_3^c\mp\sqrt{(\alpha_3^p-\alpha_3^c)^2+4(\alpha_3^{pc})^2}\right)v_R^2,
\end{eqnarray}
with
\begin{eqnarray}
&&\phi_2=\cos\xi\; H_1 - \sin\xi\; H_2,\;\;\; \chi_2=\sin\xi\; H_1 + \cos\xi\; H_2,\nonumber\\
&&\tan2\xi=(\alpha_3^p-\alpha_3^c)/\alpha_3^{pc}.
\end{eqnarray}
Mixings inside doublets $H_1$ or $H_2$ are generally $O(k^2/v_R^2)$.

\item  Heavy right-handed triplet Higgs (neutral) from Re$(\delta_R^0)$ at the $v_R$ scale:

\begin{eqnarray}
(M_R^0)^{2}&\simeq&2\rho_1v_R^2
\end{eqnarray}

\item Heavy doubly charged right-handed triplet Higgs($\delta^{\pm\pm}_R$):

\begin{eqnarray}
(M_R^{\pm\pm})^{2}&\simeq&2\rho_2v_R^2
\end{eqnarray}

\item heavy left-handed triplet $\Delta_L$ (neutral, charged and doubly charged):

\begin{eqnarray}
(M^{\Delta_L})^{2}&\simeq&\frac{1}{2}(\rho_3-2\rho_1)v_R^2
\end{eqnarray}

\item Higgs mixing among different components:

The light Higgs $h^0_{1,2,3}$ mixings with heavy Higgs Re$(\delta_R^0)$  are of order $O(\frac{k}{v_R})$.
The light Higgs $h^0_{1,2,3}$ mixings with heavy Higgs $H_{1,2}^0$ are of order $O(\frac{k^2}{v_R^2})$.
The light Higgs $h^{\pm}$ mixings with $H_{1,2}^{\pm}$ are of order $O(\frac{k^2}{v_R^2})$.
The heavy Higgs $H_{1,2}^0$ mixings with Re$(\delta_R^0)$ are of order $O(\frac{k^2}{v_R^2})$. $\Delta_L$ mixings with others approach to vanishing when $v_L\rightarrow 0$.
\end{itemize}

\end{document}